\documentclass[smallextended]{svjour3}

\usepackage{mathtools,amsmath,amssymb}
\usepackage{microtype}
\usepackage{algorithm,algpseudocode}
\usepackage{tabu,booktabs}
\usepackage[compress,numbers]{natbib}
\usepackage{hyperref}

\title{x3: Lossless Data Compressor}

\author{David Barina}

\institute{Brno University of Technology\\
Faculty of Information Technology\\
Centre of Excellence IT4Innovations\\
Brno, Czech Republic\\
\email{ibarina@fit.vutbr.cz}
}

\date{Received: date / Accepted: date}

\begin{document}

\maketitle

\begin{abstract}
x3 is a lossless optimizing dictionary-based data compressor.
The algorithm uses a combination of a dictionary, context modeling, and arithmetic coding.
Optimization adds the ability to find the most appropriate parameters for each file.
Even without optimization, x3 can compress data with a compression ratio comparable to the best dictionary compression methods like LZMA, zstd, or Brotli.
\end{abstract}

\section{Introduction}
\label{sec:Introduction}

This article is intended to provide a detailed description of the x3 compression algorithm.
The x3 is an open-source\footnote{\url{https://github.com/xbarin02/x3-compressor}} lossless data compressor based on the concept of dictionary compression methods.
The main goal in x3 design was to achieve maximal compression ratio, even at the cost of large memory requirements and long compression times.
The core of the algorithm is a dictionary.
Unlike methods based on LZ77, the x3 algorithm looks for suitable fragments in a search window, which it then explicitly inserts into the dictionary.
Each fragment stored in the dictionary gets a unique number (index), and a sequence of these indexes then forms the compressed stream.
The backend of the x3 algorithm consists of a context arithmetic encoder.

The x3 offers the possibility of optimization (search for suitable parameters for each file).
This optimization only affects the compression process and does not involve the need to transmit additional data to the decoding side.
The compressor's fundamental parameters include the size of the search window and the maximum number of matches in this window.

This article further compares the x3 algorithm with other dictionary methods used in practice.
The well-known Silesia corpus is used for this comparison.
Experimental evaluation shows that x3 can compress files with a higher compression ratio than all other state-of-the-art dictionary methods (including Brotli, zstd, and LZMA).

The rest of the article is organized as follows.
Section \ref{sec:Related Work} reviews dictionary-based compression techniques such as the LZ77 and LZ78.
Section \ref{sec:Overview of x3} provides a high-level as well as a detailed description of the algorithmic structure of the x3 compressor.
Section \ref{sec:Evaluation} provides evaluation on well-known Silesia corpus.
Finally, Section \ref{sec:Conclusions} concludes the paper.

\section{Related Work}
\label{sec:Related Work}

This section describes dictionary compression methods, especially the LZ77 and LZ78.
Their description is by no means detailed.
For readers who need to study the topic in more depth, we recommend books \cite{Salomon2007,Salomon2010} by David Salomon.

% dictionary methods
Dictionary methods parse input stream into variable-length fragments, place them into a dictionary, and then encode the fragments one by one using short tags (usually containing indexes into the dictionary).
The dictionary, therefore, holds uncompressed strings of symbols.
Since the fragments must be initially transferred in uncompressed form, the output stream contains tags (indexes) and raw fragments.
The very last step---tag encoding---often employs some statistical compression method like Huffman coding.
The compression--decompression chain is usually highly asymmetric since the compressor must find good partitioning of the input stream.
In contrast, the decompressor simply follows the tags written in the output stream.

% LZ77
The two fundamental dictionary methods are LZ77 and LZ78.
The LZ77 \cite{Ziv1977} method was published in 1977 by Israeli scientists Abraham Lempel and Jacob Ziv.
The core of the method is a sliding window moving through uncompressed data.
The window is divided into two parts: the search buffer and the look-ahead buffer.
The search buffer serves as a dictionary, whereas the look-ahead buffer contains text which has not yet been processed.
The compressor tries to find the longest possible match of the look-ahead buffer in the search buffer.
After each such step, the compressor produces a tag with information about the location (offset) and length of the match found.
The tag consists explicitly of the triple (offset, length, symbol).
The offset is a position in the search buffer where a string matching the fragment at the beginning of the look-ahead buffer was found.
The length element is the length of this string, and the symbol is the next symbol in the look-ahead buffer, which no longer matches the fragment in the search buffer.

% DEFLATE, Brotli, zstd, LZMA
Many commonly used methods are based on the concept of LZ77 (e.g., DEFLATE, LZ4, zstd, Brotli, or LZMA).
For example, DEFLATE is a combination of LZ77 and Huffman coding.
Unlike the original LZ77, the tags consist only of two elements (offset, length).
The missing element (symbol) is written to the output stream separately.
These entities are coded using two Huffman tables (one for symbols and lengths, the other for offsets).
The size of the search buffer can be up to 32 kilobytes.
Unlike the original LZ77, the compressor can defer the match selection and encode the first symbol of the look-ahead buffer separately.
Zstd and Brotli \cite{Alakuijala2018} extend the basic LZ77 principle in that they allow a sequence of symbols to be encoded at once (instead of writing each symbol separately).
Tags in LZMA (xz format) are even more diverse (e.g., a tag indicating that the offset is equal to the last used offset).
However, all these methods have one thing in common: the dictionary is a search buffer referenced by offsets and lengths.

% LZ78 and LZW
The LZ78 \cite{Ziv1978} method was published by A. Lempel and J. Ziv in 1978.
Unlike the LZ77, it is no longer based on a sliding window.
It uses a dictionary in which an integer index references fragments of the uncompressed stream.
The encoder creates tags of the form (index, symbol).
The compressor searches for the longest matching string in the dictionary and generates a tag with its index and with the next symbol that breaks the match.
At the same time, each tag specifies a new string to be placed into the dictionary.
The method has many modifications, such as the LZW algorithm.
LZW \cite{Welch1984} is a variant of LZ78, developed in 1984 by American scientist Terry Welch.
As with the LZ78, the LZW is based on a dictionary that maps index to uncompressed fragments.
However, as a result of an initialized dictionary, the tag consists of only one element (index).

As can be understood from the text above, there is a big difference between LZ77-based and LZ78-based methods.
The latter use a dictionary that maps an integer index to a fragment of the input data stream, whereas the former uses a pair (offset, length).
The method presented in this article is more or less based on the LZ78 paradigm.
It uses two tags---index referring to a fragment in a dictionary and raw fragments of the input stream.
Details are described in the following section.

% bzip2, PAQ, PPMx
For the sake of completeness, it should be noted that not all state-of-the-art compression methods are based on dictionaries.
For example, bzip2 is based on the Burrows--Wheeler transform \cite{Burrows1994} and backed by Huffman coding.
PPM \cite{Cleary1984} methods are based on maintaining statistical contexts for variously-long sequences of the preceding symbols and backed by arithmetic coding.
PAQ methods predict one bit at a time instead of one symbol, use a neural network, context mixing algorithm, and are backed by arithmetic coding.
Also, there are many other very efficient methods (e.g., cmix), but their description is outside the scope of this article.

% an overview of x3
\section{Overview of x3}
\label{sec:Overview of x3}

The x3 encoder maintains a dictionary that maps integers to fragments of uncompressed data.
The encoder offers two main options affecting compression ratio and speed.
The first one is the size of a window in which the search for the most appropriate fragment for insertion into the dictionary is performed.
The second one is the maximum number of matches inside this window (to be explained below in the text).
It turns out to be a difficult task to choose a suitable fragment to insert into the dictionary.
As one would expect, in a certain sense, the two above options trade compression speed for compression efficiency (a short window leads to a quick search for the most suitable fragment).

The compression algorithm consists of two main phrases---input parsing and entropy coding.
These are described below.
The goal of input parsing is to find the most suitable fragment in a window starting at the current position in an uncompressed stream (everything behind this position is already compressed) so the subsequent entropy coding phase will produce the shortest possible number of bits.
Note that, usually, longer and more frequently used fragments tend to lead to a better compression ratio.
The encoder can decide at any time that it does not use the fragment just found, and instead, use a fragment previously stored in the dictionary.
All the same, the necessary condition is that the fragment in the dictionary must match the string at the beginning of the window.
This decision is made based on the length of the fragment from the dictionary and the length of the fragment just found in the window.
The above algorithm is formally described in Algorithm \ref{alg:Compression}.
It is important to note that the meaning of the window in this algorithm is fundamentally different from the search window in LZ77-type algorithms.
In LZ77-type algorithms, the search window contains information already known to the decoding side.
In the algorithm presented in this paper, the window contains data that has not yet been compressed.

\begin{algorithm}
	\caption{Compression algorithm overview.}
	\label{alg:Compression}
	\begin{algorithmic}[1]
		\State $p \gets 0$\Comment{The $p$ is a position in input stream.}
		\While{$p \not= \mathrm{EOF}$}
			\State $l_d \gets \Call{QueryDictionary}{p}$\Comment{Length of the fragment.}
			\State $l_w \gets \Call{SearchInWindow}{p}$\Comment{Length of the fragment.}
			\If{$l_d > l_w$}
				\State $\Call{EncodeDictionaryIndex}{p, l_d}$
				\State $p \gets p + l_d$
			\Else
				\State $\Call{EncodeRawFragment}{p, l_w}$
				\State $\Call{AddFragmentToDictionary}{p, l_w}$
				\State $p \gets p + l_w$
			\EndIf
		\EndWhile
	\end{algorithmic}
\end{algorithm}

% input parsing
The \textsc{QueryDictionary} function queries the dictionary and returns the length of the longest\footnote{the dictionary does not have the prefix property} fragment found (or infinity if not found).
Further, the \textsc{EncodeDictionaryIndex} emits the event (the event is encoded using arithmetic coder), which indicates the index of this fragment in the dictionary.
This procedure will be detailed below.
On the other hand, the \textsc{SearchInWindow} function searches the window for the longest most frequently repeated string starting at position $p$.
The search is controlled by a parameter ceiling the maximum number of matches (thus also affecting speed--efficiency tradeoff).
The pseudocode is given in Algorithm \ref{alg:SearchInWindow}.
The \textsc{CountOccurrences} returns the number of occurrences of a fragment of the length $l$ positioned at the position $p$ in the window ($p+1$ \ldots $p+W-1$), where $W$ is the window length.
The entire process is a little more complicated because the function must avoid choosing such matches that would break the future match of substantial length, or would break the future match with the fragment already stored in the dictionary.
This introduces additional options into the encoder (enable/disable future match detection).
Since several options control the encoder, this gives us the ability to find the most suitable settings for each compressed file individually.
The disadvantage of this approach is that it is very time-consuming.
However, one can use a setting that gives good results on most files and thus avoids searching for suitable parameters.

\begin{algorithm}
	\caption{\textsc{SearchInWindow} function.}
	\label{alg:SearchInWindow}
	\begin{algorithmic}[1]
		\Require{$M$ : maximum number of matches; $L$ : maximum match length}
		\Ensure{returns length of the best match}
		\Statex
		\Function{SearchInWindow}{$p$}
			\For{$l \gets 1$ \ldots $L$}
				\State $c_l \gets \Call{CountOccurrences}{p, l}$
			\EndFor
			\For{$m \gets M$ \ldots 1}
				\For{$l \gets L$ \ldots 1}
					\If{$c_l > m$}
						\State \Return{$l$}
					\EndIf
				\EndFor
			\EndFor
		\EndFunction
	\end{algorithmic}
\end{algorithm}

% entropy encoding
Now the \textsc{EncodeDictionaryIndex} and \textsc{EncodeRawFragment} produce events which are passed on to an adaptive 2nd order context arithmetic encoder, as described in \cite{Bodden2007}.
One of the following contexts is used: the last two indexes into the dictionary, the last index into the dictionary, or no context (encode directly the index into the dictionary).
Unlike PAQ-type methods, the algorithm does not perform any context mixing.
The decision on which context to use is based on an estimate of the compression ratio using the frequency of prior occurrence.
No parameters control this part of the compression chain.
The only notable exception is the initialization of occurrence frequencies in favor of the first two named contexts.

% evaluation
\section{Evaluation}
\label{sec:Evaluation}

The evaluation in Table \ref{tab:evaluation} was performed on the well-known Silesia corpus.
We included only significant dictionary methods in comparison.
However, it must be said that there are other state-of-the-art methods (e.g., bzip2) that, in some cases, provide more efficient compression.
It should also be said that we set all evaluated programs to an extreme compression ratios: \texttt{lz4 -9}, \texttt{gzip --best}, \texttt{xz -9 -e}, \texttt{zstd --ultra -22}, and \texttt{brotli -q 11}.
The LZ4 compressor is focused on speed, not compression ratio.
It was included in the comparison because it is currently very popular.
The xz compressor implements the LZMA method.
The x3 compressor was instructed by such parameters, which were selected by the state-space search for the 8 KiB window.
Longer windows improve the compression ratio but significantly slow down the compression time.
The decompression time does not depend on the compression parameters, because decompression simply executes the events written in the bitstream.
The comparison in Table \ref{tab:evaluation} shows that x3 and xz provide superior compression performance.
Furthermore, in all cases, x3 has better performance than LZ4 and gzip.

\begin{table}
	\caption{Compression ratio on Silesia corpus. Best results in bold.}
	\label{tab:evaluation}
	\begin{tabu} to \linewidth {X[l]|X[r]|X[r]|X[r]|X[r]|X[r]|X[r]}
	\toprule
		\rowfont[c]{\bfseries}
		File & LZ4 & gzip & xz & zstd & Brotli & x3 \\
	\midrule
		dickens &   2.2948   &   2.6461   &   3.6000   &   3.5765   &   3.6044   &   \textbf{3.7168}   \\
		mozilla &   2.3176   &   2.6966   &   \textbf{3.8292}   &   3.3769   &   3.6922   &   2.7432   \\
		mr      &   2.3472   &   2.7138   &   3.6231   &   3.2132   &   3.5317   &   \textbf{4.0364}   \\
		nci     &   9.1071   &  11.2311   &  \textbf{23.1519}   &  20.7925   &  22.0780   &  19.1103   \\
		ooffice &   1.7349   &   1.9907   &   \textbf{2.5346}   &   2.3587   &   2.4818   &   2.0668   \\
		osdb    &   2.5290   &   2.7138   &   3.5456   &   3.2855   &   3.5812   &   \textbf{3.6151}   \\
		reymont &   3.1345   &   3.6396   &   5.0374   &   4.9060   &   4.9747   &   \textbf{5.1010}   \\
		samba   &   3.5122   &   3.9950   &   \textbf{5.7778}   &   5.5267   &   5.7367   &   4.1871   \\
		sao     &   1.2639   &   1.3613   &   \textbf{1.6386}   &   1.4479   &   1.5812   &   1.5042   \\
		webster &   2.9554   &   3.4372   &   4.9540   &   4.8970   &   4.9188   &   \textbf{4.9685}   \\
		xml     &   6.9277   &   8.0709   &  12.2910   &  11.8004   &  \textbf{12.4145}   &   9.2249   \\
		x-ray   &   1.1798   &   1.4035   &   1.8868   &   1.6457   &   1.8096   &   \textbf{1.9649}   \\
	\bottomrule
	\end{tabu}
\end{table}

Because the construction of internal data structures is generally memory intensive, we also compared the x3 memory consumption on Silesia corpus.
Results are given in Table \ref{tab:memory}.
Size column indicates the file size.
MaxRSS\footnote{maximum resident set size} indicates the maximum amount of space of physical memory (RAM) held by the x3 process.
Factor is the ratio MaxRSS / Size, indicating how memory intensive the compression of the file was.
The x3 compressor has been instructed with such settings that it gives good results on most files (no optimization was used).
Note that memory consumption of x3 compressor includes the entire input stream, internal data structures, and output stream.
The internal data structures consist mainly of the dictionary, first-order context structure, a tree for mapping second-order contexts to integers, and second-order context structure.
It is clear from Table \ref{tab:memory} that the memory requirements are not directly proportional to the size of the input data.
Instead, they reflect the internal structure of the compressed data.
On the Silesia corpus, the expansive factor ranges from 1.7 to 18.

\begin{table}
	\caption{Memory consumption on Silesia corpus. Size and MaxRSS are given in megabytes. Factor is the ratio MaxRSS / Size.}
	\label{tab:memory}
	\begin{tabu} to \linewidth {X[l]|X[r]|X[r]|X[r]}
	\toprule
		\rowfont[c]{\bfseries}
		File    & Size & MaxRSS & Factor \\
	\midrule
		dickens &  9.8 &  42.3 &  4.4 \\ %  43328 * 1024 / 10192446 = 4.35301516437
		mozilla & 49.0 & 697.2 & 14.3 \\ % 713904 * 1024 / 51220480 = 14.2723710516
		mr      &  9.6 &  53.1 &  5.6 \\ %  54400 * 1024 /  9970564 = 5.5870059106
		nci     & 32.0 &  53.3 &  1.7 \\ %  54592 * 1024 / 33553445 = 1.66606463211
		ooffice &  5.9 & 105.8 & 18.0 \\ % 108324 * 1024 /  6152192 = 18.0299600533
		osdb    &  9.7 &  51.2 &  5.3 \\ %  52388 * 1024 / 10085684 = 5.31895625522
		reymont &  6.4 &  27.9 &  4.4 \\ %  28548 * 1024 /  6627202 = 4.41108510047
		samba   & 21.0 & 163.6 &  7.9 \\ % 167568 * 1024 / 21606400 = 7.94161137441
		sao     &  7.0 & 100.7 & 14.6 \\ % 103096 * 1024 /  7251944 = 14.5575178187
		webster & 40.0 & 177.1 &  4.5 \\ % 181320 * 1024 / 41458703 = 4.47847295175
		xml     &  5.1 &  22.1 &  4.3 \\ %  22616 * 1024 /  5345280 = 4.33256704981
		x-ray   &  8.1 &  59.0 &  7.3 \\ %  60452 * 1024 /  8474240 = 7.30482591949
	\bottomrule
	\end{tabu}
\end{table}

Since the window size significantly affects compression performance, we decided to demonstrate the effect of this parameter on the selected file.
We have chosen the dickens file for this demonstration because it is small enough for experiments with a very long window.
The file is a simple text (a concatenation of some of Charles Dickens's works).
For each window size, we then determined the most appropriate number of matches in the window.
The future match detection parameter was disabled for simplicity.
Table \ref{tab:window} shows the selected window sizes, the optimal number of matches for each size, and the achieved compression ratio.
A longer window leads to a higher optimal number of matches as well as a higher compression ratio.
Based on our experiments, we chose an 8 KiB window as the default one (gives sufficiently good results on most files).

\begin{table}
	\caption{Impact of window size. Window size is given in kilobytes. Matches indicate the optimal number of matches for the given window size.}
	\label{tab:window}
	\begin{tabu} to \linewidth {X[r]|X[r]|X[r]}
	\toprule
		\rowfont[c]{\bfseries}
		Window size & Matches & Compression ratio \\
	\midrule
		 1 &   7 & 3.5359 \\ % 3.535896
		 2 &   9 & 3.5548 \\ % 3.554838
		 4 &  15 & 3.5684 \\ % 3.568374
		 8 &  28 & 3.5799 \\ % 3.579945
		16 &  38 & 3.5963 \\ % 3.596295
		32 &  73 & 3.6117 \\ % 3.611745
		64 & 136 & 3.6358 \\ % 3.635755
	\bottomrule
	\end{tabu}
\end{table}

\section{Conclusions}
\label{sec:Conclusions}

This article introduced x3---a new open-source dictionary compressor.
x3 demonstrates that explicit mapping of fragments to indexes coupled with a compressed stream formed by a sequence of these indexes is comparable in compression to the best currently available dictionary compression methods.
This result comes at the cost of higher memory consumption.
Unlike other existing compression methods, x3 compression can be optimized for each specific file.
This optimization substantially improves the compression ratio.
However, even without optimization, x3 can compress data with a compression ratio comparable to the state-of-the-art dictionary methods.

\subsection*{Acknowledgement}

This work was supported by The Ministry of Education, Youth and Sports of the Czech Republic from the National Programme of Sustainability (NPU II); project IT4Innovations excellence in science -- LQ1602.

\bibliographystyle{spbasic}
\bibliography{sources}

\end{document}